\documentclass[aps,pra,twocolumn,floatfix,10pt]{revtex4-1}
\usepackage[utf8]{inputenc}
\usepackage{framed}
\usepackage{graphicx}
\usepackage{amsmath}
\usepackage{lmodern}
\usepackage{epsfig}
\usepackage{helvet}
\usepackage{framed}
\usepackage{bbold}
\usepackage{subfigure}
\usepackage{hyperref}

\hypersetup{
  colorlinks   = true, %Colours links instead of ugly boxes
  urlcolor     = blue, %Colour for external hyperlinks
  linkcolor    = blue, %Colour of internal links
  citecolor   = red, %Colour of citations
}

\newcommand{\bra}[1]{\mbox{$\langle #1 |$}}
\newcommand{\ket}[1]{\mbox{$| #1 \rangle$}}
\newcommand{\braket}[2]{\mbox{$\langle #1 | #2 \rangle$}}

\def\CFT{\mbox{\tiny CFT}}

\allowdisplaybreaks[3]
\def\letter{paper} 
\def\appendix{Appendix }

\begin{document}

\title{Tensor networks as conformal transformations}
\author{Ashley Milsted}
%\email{amilsted@perimeterinstitute.ca}
\author{Guifre Vidal}
\affiliation{Perimeter Institute for Theoretical Physics, Waterloo, Ontario N2L 2Y5, Canada}  \date{\today}

\begin{abstract}
Tensor networks are often used to accurately represent ground states of quantum spin chains. Two popular choices of such tensor network representations can be seen to implement linear maps that correspond, respectively, to euclidean time evolution and to global scale transformations. In this \letter, by exploiting the local structure of the tensor networks, we explain how to also implement \textit{local} or \textit{non-uniform} versions of both euclidean time evolution and scale transformations. We demonstrate our proposal with a critical quantum spin chain on a finite circle, where the low energy physics is described by a conformal field theory (CFT), and where non-uniform euclidean time evolution and local scale transformations are conformal transformations acting on the Hilbert space of the CFT. We numerically show, for the critical quantum Ising chain, that the proposed tensor networks indeed transform the low energy states of the periodic spin chain in the same way as the corresponding conformal transformations do in the CFT.
\end{abstract}

\maketitle
\begin{figure}
\includegraphics[width=\linewidth]{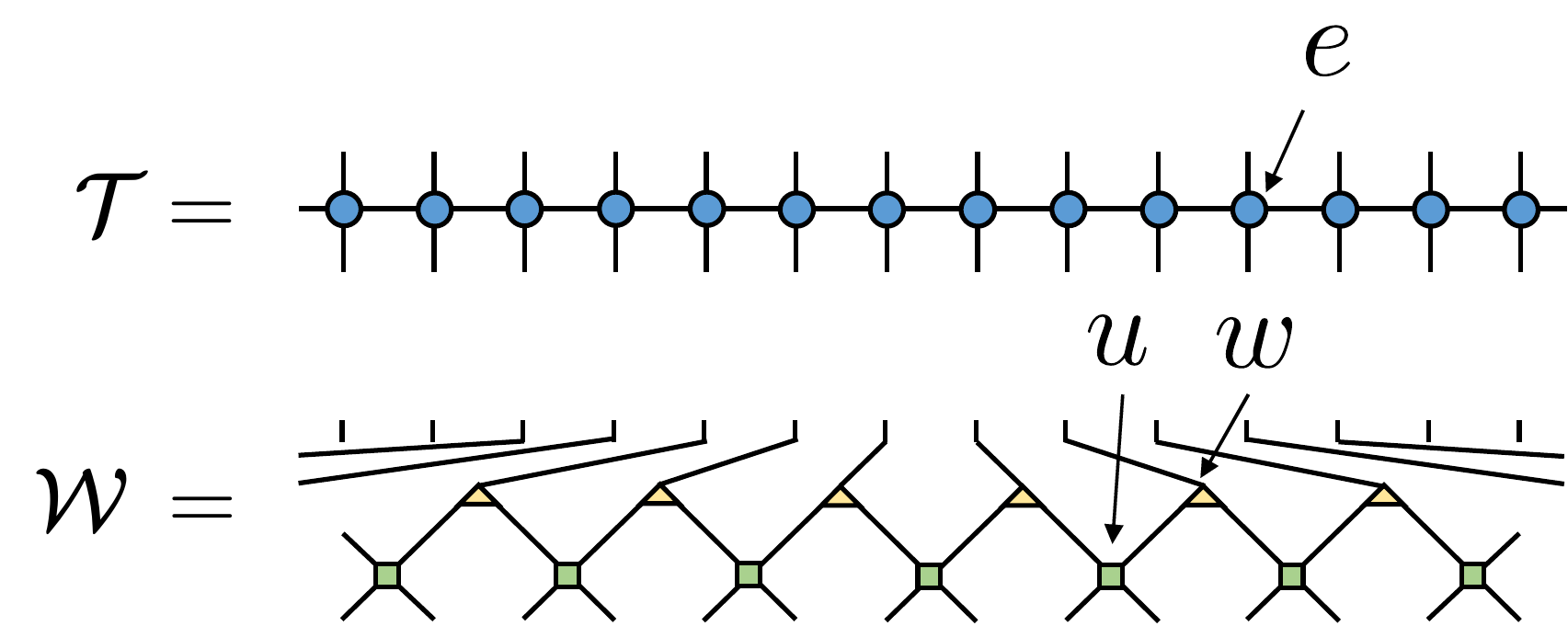}
\caption{
Tensor networks $\mathcal{T}$ and $\mathcal{W}$ for an euclidean time evolution $e^{-H}$ by unit time, made of euclideons $e$, and a scale transformation by a scale factor $2$, made of disentanglers $u$ and isometries $w$.
\label{fig:MTW} 
}
\end{figure}

Tensor networks offer a natural framework to describe complex objects in a growing number of fields, most notably in condensed matter \cite{MPS,MERA,PEPS}, but also in statistical mechanics \cite{statistical,TNR}, quantum chemistry \cite{QChem}, holography \cite{H2,dS2,holo1,holo2,holo3}, quantum and classical information theory \cite{holo3,QErrorC}, and machine learning \cite{MachineLearning}.
The main application of tensor networks to date is as an ansatz to efficiently represent low energy  states of a local quantum Hamiltonian $H$ on the lattice. For instance, the \textit{multi-scale entanglement renormalization ansatz} (MERA) can be used to accurately represent the ground state of a critical quantum spin chain \cite{MERA}. However, tensor networks can also encode linear maps between quantum many-body states. Two prominent examples, depicted in Fig.~\ref{fig:MTW}, are a tensor network $\mathcal{T}$ implementing \textit{euclidean time evolution} and a double layer $\mathcal{W}$ of MERA tensors enacting a \textit{scale transformation}. In this \letter\ we will exploit the inherent local structure of $\mathcal{T}$ and $\mathcal{W}$ to design new tensor networks that implement local versions of both euclidean time evolution and scale transformations. Our construction is of interest for several reasons. On the one hand it illustrates how conformal symmetry, which is manifestly broken by the lattice, re-emerges in the low energy physics of the critical spin chain. In addition, understanding the linear map that a tensor network implements can be used to endow that tensor network with a precise geometric meaning \cite{PathIntegral,geoMERA}, thus contributing to current attempts to use tensor networks to describe quantum gravity \cite{H2,dS2,holo1,holo2,holo3}. Finally, our work forms the theoretical basis for a novel non-perturbative approach to interacting quantum field theories in curved spacetimes \cite{QFTCS}.

\begin{figure}
\includegraphics[width=\linewidth]{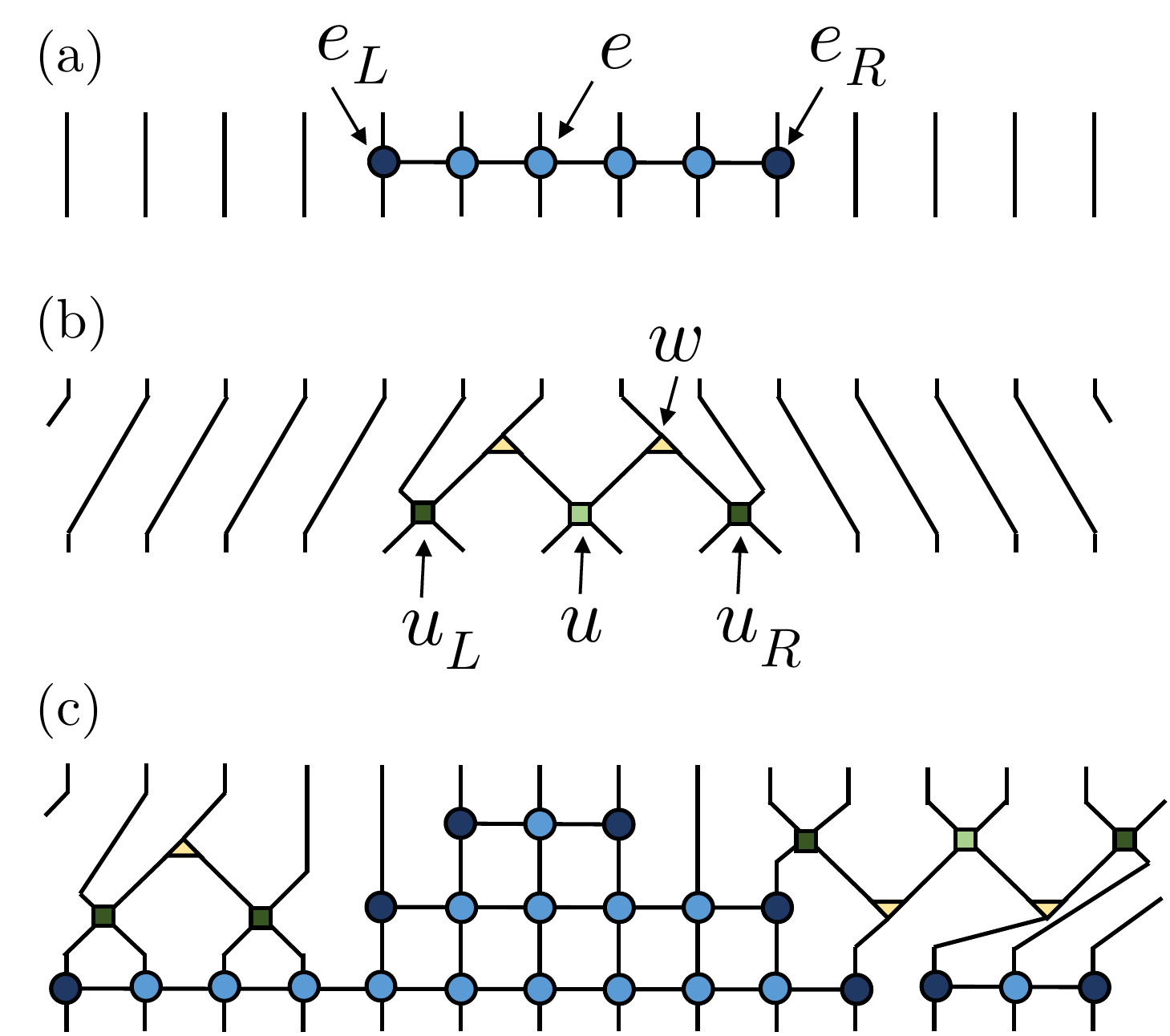}
\caption{
Examples of proposed tensor networks for (a) non-uniform euclidean time evolution, made of euclideons $e$ and smoothers $e_L$ and $e_R$, and (b) local scale transformation, made of disentanglers $u$ and isometries $w$, as well as smoothers $u_L$ and $u_R$. (c) A more general conformal transformation mixing non-uniform euclidean time evolution and rescaling is possible by combining the above building blocks. 
\label{fig:local} 
}
\end{figure}

As in several previous contributions \cite{H2, dS2, holo1, TNRMERA, TNRscale, Quotient, TakayanagiTNR, CzechTNR}, we consider a specific class of tensor networks, made mostly of \textit{euclideons} $e$ -- the tensors in the euclidean time evolution map $\mathcal{T}$ -- and of \textit{disentanglers} $u$ and \textit{isometries} $w$ -- the tensors in the MERA rescaling map $\mathcal{W}$. The main intuition is that if we apply a truncated version of $\mathcal{T}$ or $\mathcal{W}$ locally on a region of the spin chain while leaving the rest of the chain untouched, see Fig.~\ref{fig:local}, then we should be able to evolve in euclidean time, or rescale, just the spins in that region. This simple intuition turns out to be correct provided that we place additional tensors called \textit{smoothers} at both ends of the truncated network. In order to show that the proposed tensor networks implement the intended linear maps, we consider their action on the Hilbert space of a periodic critical quantum spin chain whose low energy physics is described by a conformal field theory (CFT) \cite{CFT1,CFT2}. With the critical quantum Ising model as a concrete example, we first numerically diagonalize the lattice Hamiltonian $H$ at low energies and then use the techniques of Refs.~\cite{Cardy,KooSaleur,Ash} to match individual low energy states of the critical spin chain with their CFT counterparts. In this way we e.g.\ identify the \textit{conformal towers}-- that is, the irreducible representations of the conformal group. Finally, we apply the proposed tensor network linear maps, which should implement specific \textit{conformal transformations}, and numerically confirm that they indeed act diagonally in the conformal towers and transform individual low energy states within each tower as the expected non-uniform euclidean time evolution and rescaling.

\textit{Tensor networks for euclidean time evolution and rescaling.---} Consider a critical quantum spin chain on the circle made of $N$ spins and with Hamiltonian $H$. As linear maps, the two tensor networks $\mathcal{T}$ and $\mathcal{M}$ of Fig.~\ref{fig:MTW} act on the Hilbert space of the spin chain in a well-understood way \cite{MERA,TNR}. The transfer matrix $\mathcal{T}$, made of $N$ euclideons, implements a uniform euclidean time evolution $e^{-H}$ of the $N$ spins, whereas the network $\mathcal{W}$, made of $N$ disentanglers $u$ and $N$ isometries $w$, implements a scale transformation by a scale factor $2$ that embeds the Hilbert space of the chain of $N$ spins into that of a new chain made instead of $2N$ spins. In these constructions, the euclideon $e$ can be obtained from the Hamiltonian $H$, whereas tensors $u$ and $w$ are extracted from a MERA that approximates the ground state of $H$ \cite{SupplMat}. We can then accurately represent this ground state either by concatenating many copies of the transfer matrix $\mathcal{T}$ (which implement a time evolution by a large euclidean time) or many copies of the double layered $\mathcal{W}$ (resulting in the MERA). As a matter of fact, these two tensor network representations are not independent, but related by a coarse-graining algorithm, known as \textit{tensor network renormalization} \cite{TNR}, that transforms euclideons $e$ into optimized disentanglers $u$ and isometries $w$ \cite{TNRMERA}.

The local structure of the tensor network $\mathcal{T}$ strongly suggests that we may be able to use a truncated row of euclideons $e$ acting only on a region of the spin chain as a means to apply a time evolution only on that region, see Fig.~\ref{fig:local}(a). Notice that our proposal includes two new tensors, \textit{smoothers} $e_L$ and $e_R$ (obtained by properly splitting an euclideon $e$ \cite{SupplMat}), which are placed at the left and right ends of the truncated row of euclideons in order to eliminate dangling bond indexes. 
Similarly, we also propose the use of a truncated $\mathcal{W}$ of disentanglers $u$ and isometries $w$, Fig.~\ref{fig:local}(b), as a means to apply a local rescaling by a factor 2 only on a certain region of the spin chain. Again, smoothers $u_L$ and $u_R$ (obtained from $u$ and $w$ \cite{SupplMat}) are used at both ends of the truncated $\mathcal{W}$. A more general transformation mixing non-uniform euclidean time evolution and local rescaling, see Fig.~\ref{fig:local}(c) for an example, can then be obtained by combining truncated rows of euclideons $e$ and/or of disentanglers $u$ and isometries $w$ together with appropriate smoothers $e_{L}, e_{R}, u_{L}, u_{R}$.

\textit{Low energy states and conformal symmetry.---} To characterize the action of the proposed tensor networks as linear maps, we will apply them to the low energy states of a critical quantum spin chain on a finite circle. Let $\ket{\phi_{\alpha}}$ denote the simultaneous eigenvectors of the lattice Hamiltonian $H$ and one-site translation operator $T$,
\begin{equation}
H \ket{\phi_{\alpha}} = E_{\alpha} \ket{\phi_{\alpha}},~~~~T\ket{\phi_{\alpha}} = e^{-iP_{\alpha}} \ket{\phi_{\alpha}},
\end{equation}
where $E_{\alpha}$ and $P_{\alpha}$ denote the energy and momentum of $\ket{\phi_{\alpha}}$. Using (\textit{i}) Cardy's expression relating the energy and momentum $(E_\alpha,P_\alpha)$ of \textit{low energy} lattice states $\ket{\phi_{\alpha}}$ to the scaling dimension and conformal spin $(\Delta_\alpha,S_\alpha)$ of \textit{scaling operators} $\phi_{\alpha}$ of the continuum CFT \cite{Cardy}, and (\textit{ii}) the Koo-Saleur formula relating the local terms in the lattice Hamiltonian $H$ to the Virasoro generators $L_n, \bar{L}_n$ of the conformal group \cite{KooSaleur}, Ref.~\cite{Ash} recently explained how to match individual low energy states $\ket{\phi_{\alpha}}$ with CFT scaling operators $\phi_{\alpha}$ and e.g.\ organize them into conformal towers (irreducible representations of the conformal group \cite{CFT2}). More generally, (\textit{i}) and (\textit{ii}) tell us how conformal transformations (including non-uniform euclidean time evolution and rescaling) should act on the low energy states $\ket{\phi_{\alpha}}$. This allows for direct comparison between numerical results on a periodic chain made of $N$ spins and theoretical predictions from a CFT on a circle of circumference $L=N$. For instance, the transfer matrix $\mathcal{T}$ and the lattice translation operator $T$ can be seen to act on low energy states as $\mathcal{T} \sim e^{-H^{\mbox{\CFT}}}$ and $T \sim e^{-iP^{\mbox{\CFT}}}$, where 
\begin{equation} \label{eq:HP}
H^{\mbox{\tiny CFT}} \equiv \int_0^{L} dx~h(x),~~~P^{\mbox{\tiny CFT}}\equiv \int_0^{L} dx~p(x),
\end{equation}
are the CFT hamiltonian and momentum operators, expressed in terms of the CFT hamiltonian and momentum densities $h(x)$ and $p(x)$. In turn, the tensor network $\mathcal{W}$ is seen to act on low energy states as the identity, as further analysed in Ref.~\cite{geoMERA}.

\begin{figure}
\includegraphics[width=\linewidth]{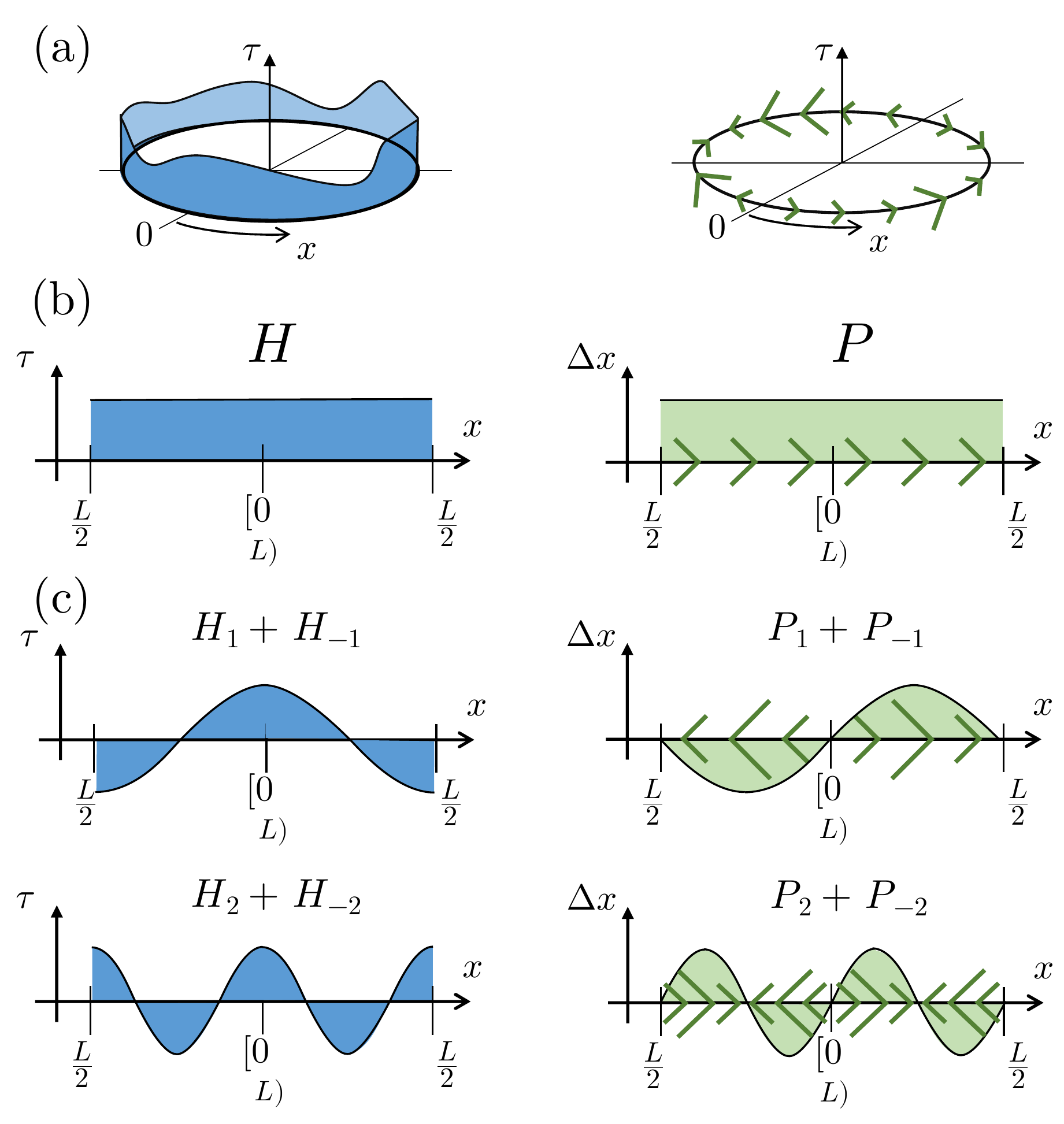}
\caption{ (a) When regarding the circle as embedded in a spacetime cylinder at euclidean time $\tau=0$, we can consider two types of deformations of the circle by conformal transformations: time evolution (left), which displaces the points in the circle in the time direction, and deformations of the circle at constant time (right), which map points of the circle into other points of the same circle. 
(b) Uniform conformal deformations include time evolution and a periodic translation, as generated by the hamiltonian $H^{\CFT}$ and the momentum operator $P^{\CFT}$ in Eq.~(\ref{eq:HP}). 
(c) Non-uniform deformations are generated by $Q_0$ and $Q_1$ in Eqs.~(\ref{eq:Q0})-(\ref{eq:Q1}), which can be expanded in Fourier modes $H_n$ and $P_n$, shown here for $n=1,2$. 
\label{fig:deform} 
}
\end{figure}

\textit{Deformations of the circle.---} In a CFT, infinitesimal conformal transformations are naturally organized into non-uniform time evolution and rescaling on the circle, see Fig.~\ref{fig:deform}. A conformal transformation $V_0^{\CFT} = e^{-Q_0}$ corresponding to non-uniform time evolution is generated by \cite{Ash}
\begin{equation} \label{eq:Q0}
Q_0 \equiv \int_0^{L} dx~a(x)~h(x) = \frac{2\pi}{L}\sum_{n\in\mathbb{Z}} a_n H_n, 
\end{equation}
where the operators $H_n \equiv L_n + \bar{L}_{-n} - (c/12) \delta_{n,0}$ are the Fourier modes of the CFT hamiltonian density $h(x)$ \cite{SupplMat}. Similarly, a conformal transformation $V_1^{\CFT} = e^{-iQ_1}$ corresponding to a non-uniform rescaling of the circle into itself is generated by 
\begin{equation} \label{eq:Q1}
Q_1 \equiv \int dx~b(x)~p(x) = \frac{2\pi}{L}\sum_{n\in\mathbb{Z}} b_n P_n,
\end{equation}
where the operators $P_n \equiv L_n - \bar{L}_{-n}$ are the Fourier modes of the CFT momentum density $p(x)$. The profile functions $a(x)$, $b(x)$ are two real functions on the circle $[0,L)$ which indicate displacements in time $\tau(x)$ and position $\Delta x(x)$, whereas $a_n$, $b_n$ are their Fourier components \cite{SupplMat}.

\begin{figure}
  \includegraphics[width=\linewidth]{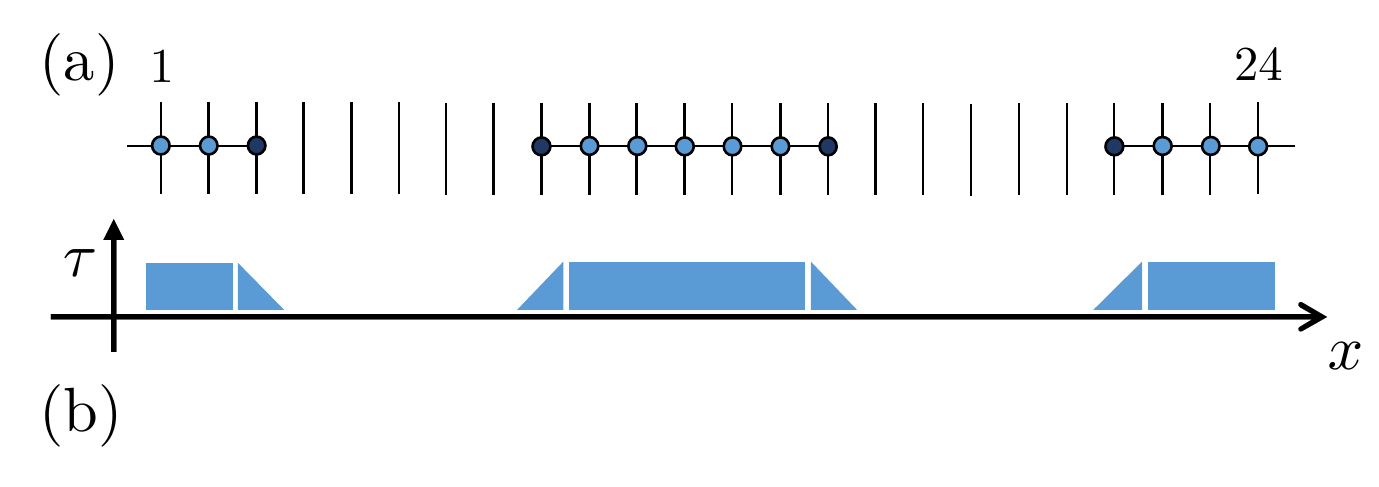}\\
  \vspace{-0.2cm}
  \includegraphics[width=0.95\linewidth]{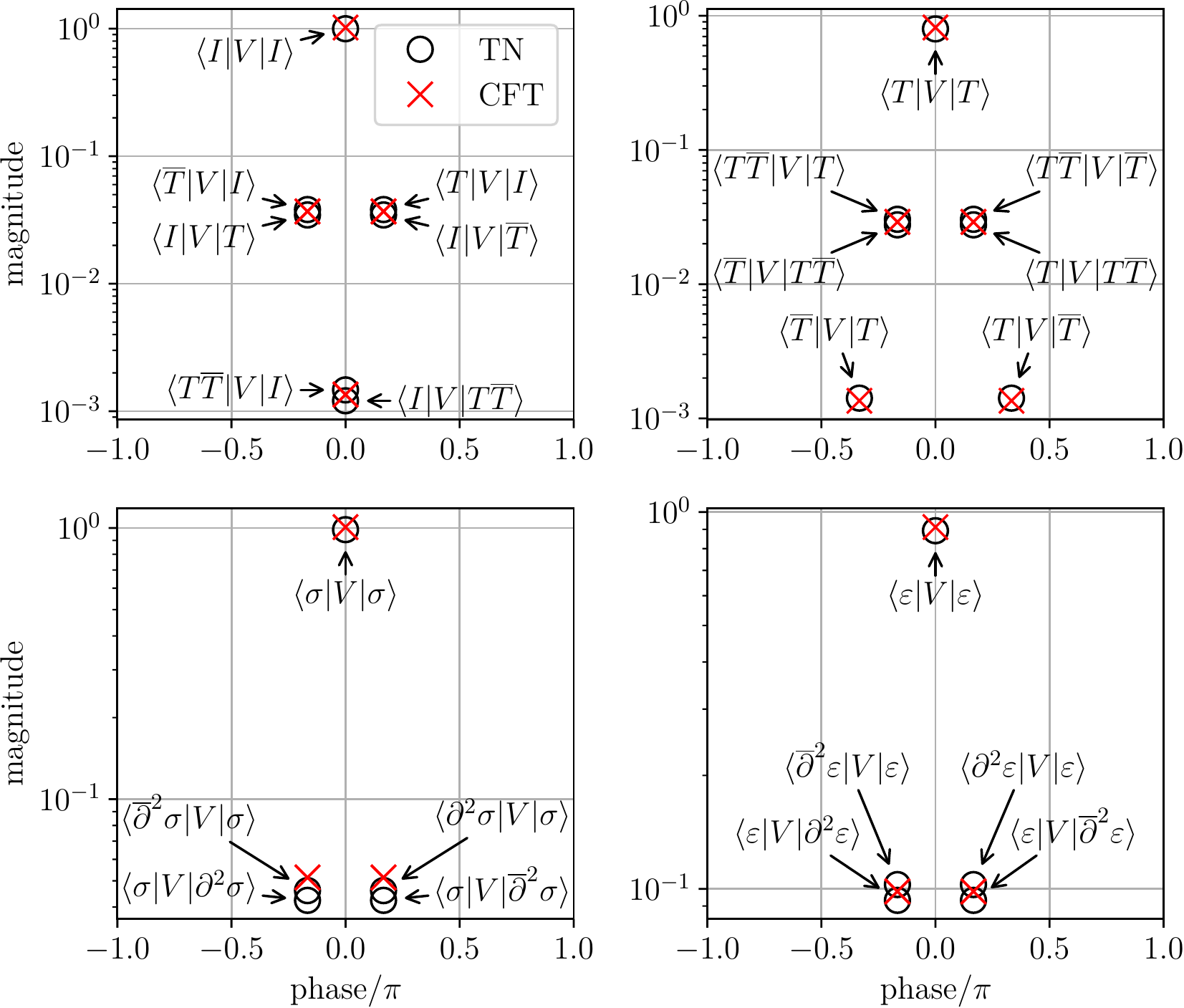}
\caption{
(a) Tensor network for a non-uniform euclidean time evolution $V$ on $N=24$ spins together with the euclidean time evolution profile which it is expected to approximate \cite{SupplMat}.
(b) Result of applying $V$ to the ground state $\ket{\mathbb{I}}$ (i.e. the identity primary state), the stress-tensor states $|T\rangle$, $|\overline{T}\rangle$, and the primary states $|\sigma\rangle$ and $|\varepsilon\rangle$. 
Shown are the largest matrix elements $V_{\alpha\beta}$, see Eq.~(\ref{eq:amplitudes}), with other states in the same conformal tower as the initial state. Also shown are the corresponding matrix elements, computed perturbatively, of a conformal transformation $V_0^{\CFT}$ in the Ising CFT \cite{SupplMat}. Spurious tower-mixing matrix elements of $V$ (not shown) were found to have a maximum magnitude of $10^{-2}$ and to decrease with system size \cite{PathIntegral}.
\label{fig:time} 
}
\end{figure}

\begin{figure}
  \includegraphics[width=\linewidth]{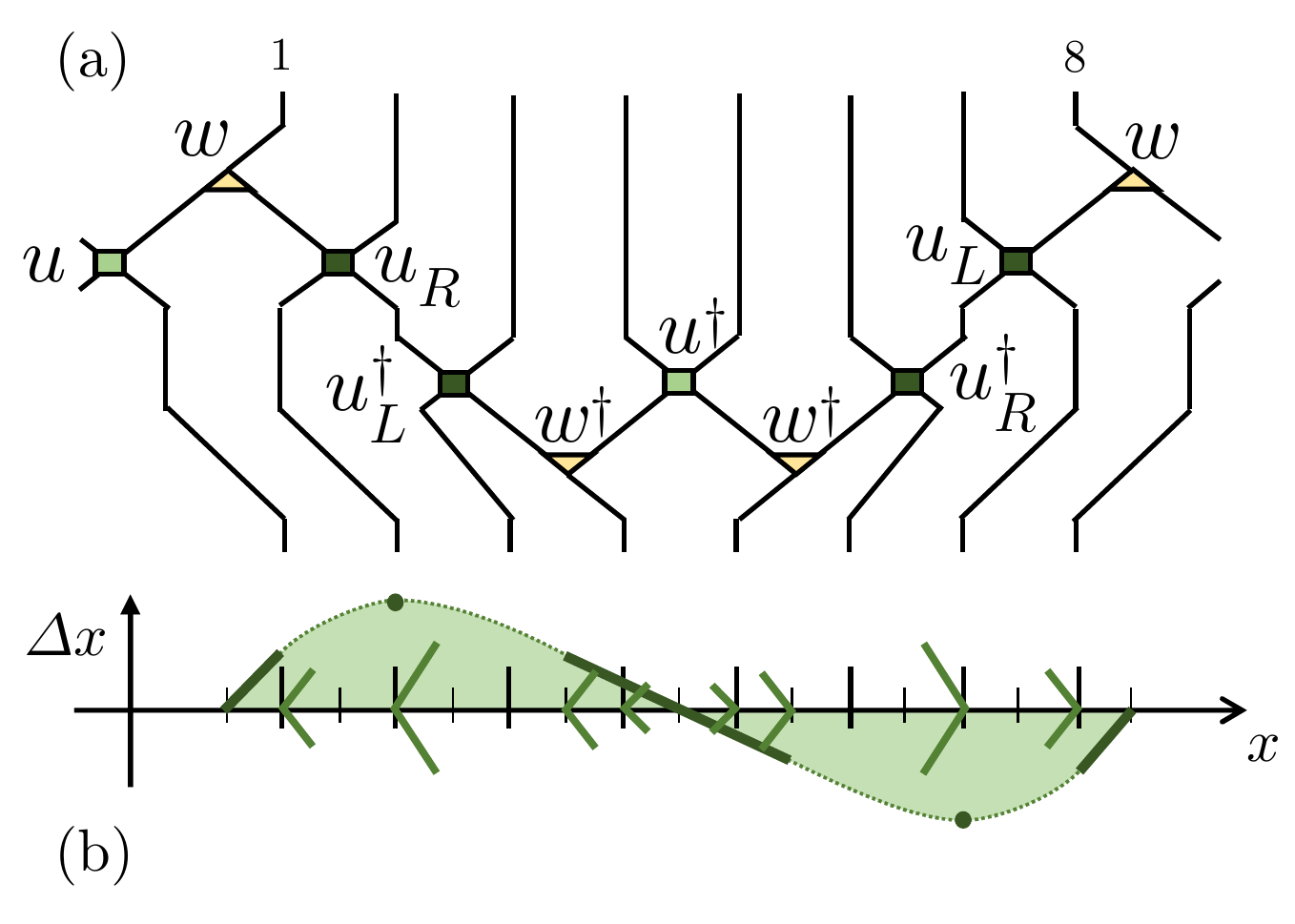} \\
  \vspace{-0.1cm}
  \includegraphics[width=0.95\linewidth]{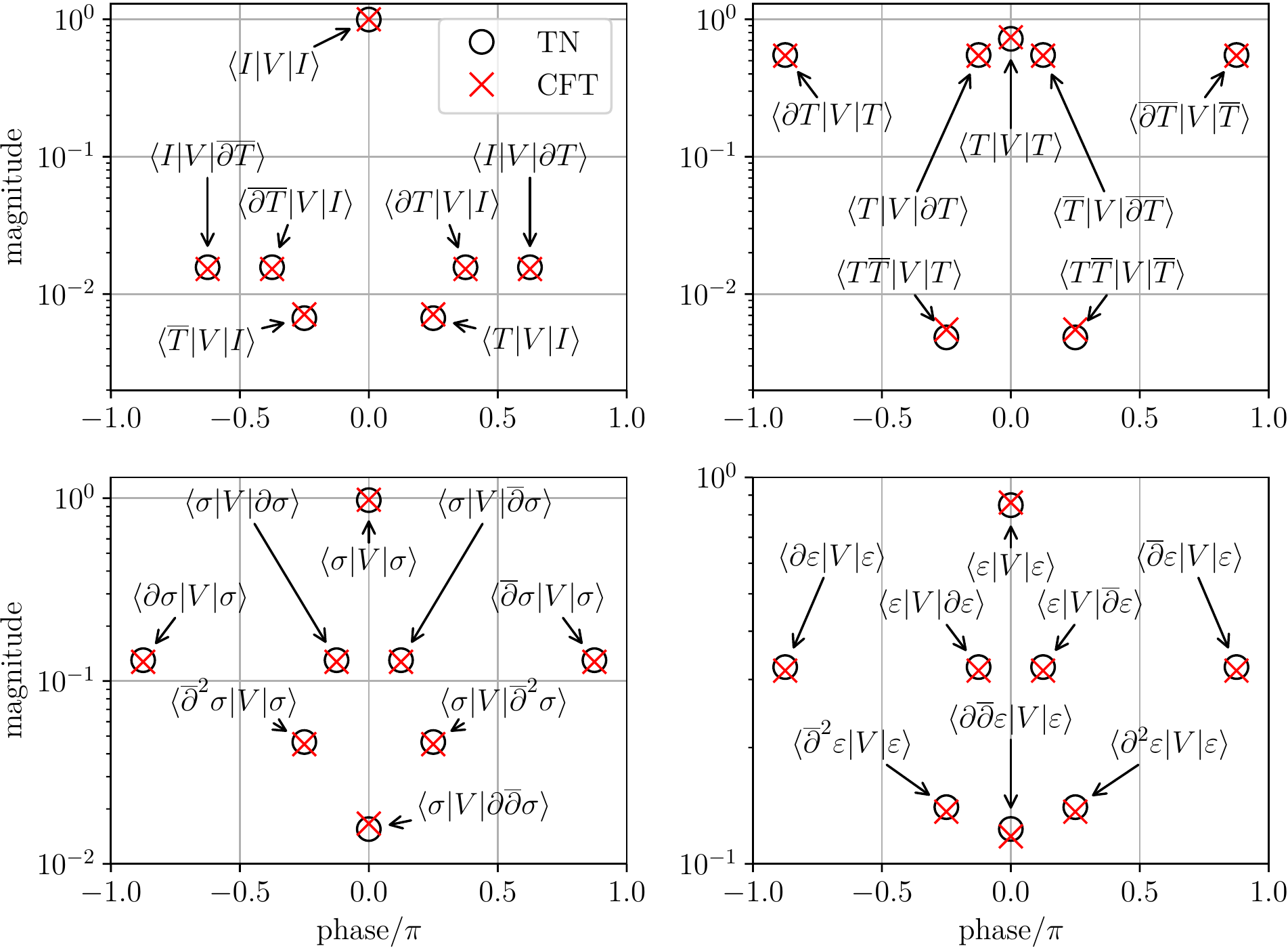}
\caption{
(a) Tensor network for a non-uniform scale transformation $V$ on a chain of $N=8$ effective sites (resulting from a scale-invariant MERA \cite{siMERA} with each effective site represented by a vector space of dimension $\chi=8$) together with the non-uniform translation profile it is expected to approximate  \cite{SupplMat}.
(b) Result of applying $V$ to the ground state $\ket{\mathbb{I}}$, the stress-tensor states $|T\rangle$, $|\overline{T}\rangle$, and the primary operator states $|\sigma\rangle$ and $|\varepsilon\rangle$. Shown are the largest matrix elements $V_{\alpha\beta}$, see Eq.~\ref{eq:amplitudes}, with other states in the same conformal tower as the initial state, as well as the corresponding matrix elements, computed perturbatively, of a conformal transformation $V_1^{\CFT}$ in the Ising CFT \cite{SupplMat}. Tower-mixing matrix elements of $V$ involving these starting states were found to have a maximum magnitude of $\sim 5\times 10^{-3}$ ($6.5 \times 10^{-4}$ for elements with $|T\rangle$ and $|\overline{T}\rangle$).
\label{fig:scale} 
}
\end{figure}

\textit{Numerical validation.---} For the critical Ising model, with Hamiltonian $H = -\sum_l \sigma^{x}_l\sigma^{x}_{l+1} - \sigma_l^{z}$, we used the techniques of Ref.~\cite{Cardy,KooSaleur,Ash} to match each low-energy eigenstate $\ket{\phi_{\alpha}}$ of $H$ and $T$ with its corresponding CFT scaling operator $\phi_{\alpha}$. In this way, the states $\ket{\phi_{\alpha}}$ were organized into three conformal towers corresponding to the three Virasoro primaries \textit{identity} $\mathbb{I}$, \textit{spin} $\sigma$, and \textit{energy density} $\varepsilon$ of the Ising CFT \cite{CFT2}. Importantly, thanks to the Koo-Saleur formula, which effectively amounts to having an approximate, lattice version of the Virasoro generators $L_n$ and $\bar{L}_{n}$, we were able to fix the otherwise random, relative complex phases between states $\ket{\phi_{\alpha}}$ in the same conformal tower that appear in diagonalizing $H$ and $T$ \cite{SupplMat}. After constructing euclideons $e$, optimized disentanglers $u$ and isometries $w$, and corresponding smoothers $e_L$, $e_R$, $u_L$ and $u_R$ for this critical spin chain, we built several tensor networks, such as those in Figs.~\ref{fig:time} and \ref{fig:scale}, which implement a linear map 
\begin{equation} \label{eq:amplitudes}
 V\ket{\phi_{\alpha}} = \sum_{\beta} V_{\alpha\beta} \ket{\phi_{\beta}},~~~~V_{\alpha\beta} \equiv  \bra{\phi_{\beta}}V \ket{\phi_{\alpha}}.
\end{equation}
In all cases we observed that, as expected of a conformal transformation, $V$ acted on low energy states approximately diagonally in the conformal towers, i.e. $V_{\alpha\beta} \approx 0$ whenever $\ket{\phi_{\alpha}}$ and $\ket{\phi_{\beta}}$ belong to different towers (with small numerical violations that are seen to decrease as we increase the size $N$ of the chain \cite{PathIntegral}). Importantly, this first non-trivial confirmation that the proposed gates implement conformal transformations required a proper choice of smoothers $e_{L},e_{R}, u_{L}, u_{R}$ \cite{SupplMat}. Moreover,  the amplitudes $V_{\alpha\beta}$ within each conformal tower were seen to also accurately match the CFT prediction for the intended conformal transformations, as shown in the two examples of Figs.~\ref{fig:time} and \ref{fig:scale}. 

\textit{Discussion.---} In this \letter\ we have proposed tensor networks that implement discrete conformal transformations on the Hilbert space of a critical quantum spin chain.  Specifically, tensor networks for non-uniform euclidean time evolution and rescaling were numerically tested and seen to reproduce the CFT predictions accurately. Remarkably, all the proposed gates are made of a small set of fixed tensors $e,u,w,e_L,e_R,u_L,u_R$ that depend on the critical spin chain but are independent of the size $N$ of the spin chain and of the specific conformal transformation being implemented. All these tensors can in turn be obtained from a single euclideon $e$ (produced either from the critical Hamiltonian $H$ or from the local Boltzmann weights of a critical classical statistical partition function for a dual 2d classical spin model \cite{SupplMat}). The construction of (a discrete set of) finite conformal transformations is a vivid manifestation of the emergence of conformal symmetry in a lattice model and complements Ref.~\cite{Ash}, which focused instead on building generators of infinitesimal conformal transformations from $H$. The present approach can also be generalized to non-critical spin chains, corresponding to massive field theories, where different tensors $e(s)$, $u(s)$, $w(s)$, etc.\ are required at each relevant length scale $s$.

The proposed tensor network constructions have a number of applications. They allow for a (numerically verified) interpretation of these tensor networks, which include those previously used in Refs.~\cite{H2, dS2, holo1, TNRMERA, TNRscale, Quotient, TakayanagiTNR, CzechTNR}, as representing an euclidean CFT path integral on a curved spacetime \cite{PathIntegral}. On the one, this is of immediate interest as a new computational framework to study interacting quantum field theory in curved spacetimes \cite{QFTCS}. On the other hand, the ability to attach a concrete geometry to the tensor network, namely that of the corresponding curved spacetime, is also of considerable interest in the context of holography, where e.g.\ MERA had been previously conjectured to realize either the hyperbolic plane \cite{H2} or de Sitter spacetime \cite{dS2}. In Ref.~\cite{geoMERA} we will propose, based on the analysis presented here, that a third, intermediate geometry with null metric might actually be more appropriate, while also introducing euclidean and lorentzian extensions of MERA that are naturally attached to the two curved geometries previously proposed in Refs.~\cite{H2} and \cite{dS2}.

%\begin{acknowledgments}
\textit{Acknowledgments.} We thank 
Bartlomiej Czech,
Pawel Caputa,
Davide Gaiotto,
Qi Hu,
Lampros Lamprou,
Juan Maldacena,
David Mateos,
Samuel McCandlish,
Rob Myers,
James Sully,
Vasudev Shyam, 
Tadashi Takayanagi, 
Xiao-liang Qi,
and, very specially, Rob Myers, for fruitful discussions and feedback.
The authors acknowledge
support by the Simons Foundation (Many Electron
Collaboration), by NSERC (discovery grant), and by Compute Canada. Research at Perimeter Institute is supported by the Government of Canada through the Department of Innovation, Science and Economic Development Canada and by the Province of Ontario through the Ministry of Research, Innovation and Science.
%\end{acknowledgments}

\section{Conformal generators on the circle}

\subsection{Virasoro generators}

A basis of generators of conformal symmetry on the circle $x \in [0,L)$ is given by the Virasoro generators $L_n$ and $\bar{L}_{n}$ \cite{CFT2}, 
\begin{eqnarray}
L_n &\equiv& \frac{L}{(2\pi)^2} \int_{0}^{L} dx~e^{i\frac{2\pi}{L}nx } T(x) + \frac{c}{24}\delta_{n,0},\\
\bar{L}_n &\equiv& \frac{L}{(2\pi)^2} \int_{0}^{L} dx~e^{-i\frac{2\pi}{L}nx } \bar{T}(x) + \frac{c}{24}\delta_{n,0},
\end{eqnarray}
which are the Fourier modes of the holomorphic and antiholomorphic components of the stress tensor
\begin{eqnarray}
T(x) &\equiv&  2\pi\frac{h(x) + p(x)}{2},\\
\bar{T}(x) &\equiv& 2\pi\frac{h(x) - p(x)}{2}.
\end{eqnarray}
Here $h(x)$ and $p(x)$ are the CFT hamiltonian and momentum densities, which we can also use to write the CFT hamiltonian and momentum operators on the circle,
\begin{eqnarray}
H^{\mbox{\tiny CFT}} &\equiv& \int_0^{L} dx~h(x) = \frac{2\pi}{L} \left(L_0 +\bar{L}_0 - \frac{c}{12}\right),\\
P^{\mbox{\tiny CFT}} &\equiv& \int_0^{L} dx~p(x) =  \frac{2\pi}{L} \left(L_0 -\bar{L}_0\right).
\end{eqnarray}

The Virasoro generators $L_n, \bar{L}_n$ can be seen to close two copies of the Virasoro algebra,
\begin{eqnarray} \label{eq:Virasoro}
\big[L_n, L_m\big] &=& (n-m)L_{n+m} + \frac{c}{12}n(n^2-1)\delta_{n+m,0},~~~~~ \\
\big[ L_n, \bar{L}_m \big] &=& 0, ~~~\\
\big[\bar{L}_n, \bar{L}_m\big] &=& (n-m)\bar{L}_{n+m} + \frac{c}{12}n(n^2-1)\delta_{n+m,0},~~~~~
\end{eqnarray} 
where $c$ is the central charge of the CFT.

\subsection{Generators of non-uniform euclidean time evolution}

Let us now consider the Fourier mode expansion of the Hamiltonian density $h(x)$,
\begin{eqnarray}
h(x) &\equiv&  \left(\frac{2\pi}{L}\right) \frac{1}{L} \sum_n H_n ~e^{-i\frac{2\pi}{L}nx},\\
H_n &\equiv& \left(\frac{L}{2\pi}\right) \int_{0}^{L}dx~ e^{i\frac{2\pi}{L}nx} ~h(x),
\end{eqnarray}
where we use the unusual factor $L/2\pi$ so that 
\begin{equation}
H_n \equiv L_{n} + \bar{L}_{-n} -\frac{c}{12}\delta_{n,0}.
\end{equation} 
In particular the Hamiltonian reads $H^{\CFT} = (2\pi/L) H_0$.
Similarly, let $a(x)$ be a real function on the circle $x \in [0,L)$, with Fourier expansion
\begin{eqnarray}
a(x) &\equiv&   \sum_n a_n ~e^{i\frac{2\pi}{L}nx},\\
a_n &\equiv& \frac{1}{L} \int_{0}^{L} dx~ e^{-i\frac{2\pi}{L}nx}~a(x).
\end{eqnarray}
Then we have that the generator $Q_0$ of a non-uniform euclidean time evolution with profile function $a(x)$ reads
\begin{eqnarray}
Q_0 &\equiv& \int_0^{L} dx~ a(x) h(x) \\
&=& \frac{2\pi}{L}\frac{1}{L}\int_0^L dx~  \sum_n a_n ~e^{i\frac{2\pi}{L}nx} \sum_m H_m ~e^{-i\frac{2\pi}{L}mx}~~~~\\
&=&  \frac{2\pi}{L}~  \sum_n a_n ~ \sum_m H_m \left(\frac{1}{L}\int_0^L dx~e^{i\frac{2\pi}{L}(n-m)x}\right)~~~~\\
&=&  \frac{2\pi}{L} \sum_{n} a_nH_n,
\end{eqnarray}
where we used that $(1/L)\int_0^L dx ~e^{i\frac{2\pi}{L}(n-m)x} = \delta_{m,n}$.

\subsection{Generators of non-uniform rescaling}

Let us now consider the Fourier mode expansion of the momentum density $p(x)$,
\begin{eqnarray}
p(x) &\equiv&  \left(\frac{2\pi}{L}\right) \frac{1}{L} \sum_n P_n ~e^{-i\frac{2\pi}{L}nx},\\
P_n &\equiv& \left(\frac{L}{2\pi}\right) \int_{0}^{L}dx~ e^{i\frac{2\pi}{L}nx} ~p(x),
\end{eqnarray}
where we again use the unusual factor $L/2\pi$ so that 
\begin{equation}
P_n \equiv L_{n} - \bar{L}_{-n},
\end{equation}
with the momentum operator reading $P^{\CFT} = (2\pi/L)P_0$. Let $b(x)$ be a real function on the circle $x \in [0,L)$, with Fourier expansion
\begin{eqnarray}
b(x) &\equiv&   \sum_n b_n ~e^{i\frac{2\pi}{L}nx},\\
b_n &\equiv& \frac{1}{L} \int_{0}^{L} dx~e^{-i\frac{2\pi}{L}nx}~b(x).
\end{eqnarray}
Then the generator $Q_1$ of a non-uniform rescaling with profile function $b(x)$ reads
\begin{eqnarray}
Q_1 \equiv \int_0^{L} dx~ b(x) p(x) =  \frac{2\pi}{L} \sum_{n} b_nP_n.
\end{eqnarray}

\section{Low energy matrix elements}
 
Consider a CFT on a circle of perimeter $L$ and a conformal transformation $V^{\CFT} \equiv e^{- Q}$, with generator $Q$, acting on its Hilbert space. Our goal is to compute the matrix elements
\begin{eqnarray}
V^{\CFT}_{\alpha \beta} &\equiv& \bra{\phi_\beta^{\CFT}} V^{\CFT} \ket{\phi_{\alpha}^{\CFT}}\\
&=& \sum_{n=0}^{\infty} \frac{(-1)^{n}}{n!}\bra{\phi_\beta^{\CFT}} Q^n \ket{\phi_{\alpha}^{\CFT}},
\end{eqnarray}
between simultaneous eigenstates of the hamiltonian and momentum operators $H^{\CFT}$ and $P^{\CFT}$,
\begin{eqnarray}
H^{\CFT} \ket{\phi_{\alpha}^{\CFT}} &=& E^{\CFT}_{\alpha} \ket{\phi_{\alpha}^{\CFT}}, \\
P^{\CFT} \ket{\phi_{\alpha}^{\CFT}} &=& P^{\CFT}_{\alpha} \ket{\phi_{\alpha}^{\CFT}},
\end{eqnarray}
where, by the CFT operator-state correspondence \cite{CFT2}, 
\begin{eqnarray}
E^{\CFT}_{\alpha} &=& \frac{2\pi}{L}\left( \Delta_{\alpha} -\frac{c}{12}\right),\\
P^{\CFT}_{\alpha} &=& \frac{2\pi}{L}S_{\alpha}.
\end{eqnarray}
Here $\Delta_{\alpha} \equiv h_{\alpha}+\bar{h}_{\alpha}$ and $S_{\alpha}\equiv h_{\alpha} -\bar{h}_{\alpha}$ are the scaling dimension and conformal spin of the scaling operator $\phi_{\alpha}$ corresponding to state $\ket{\phi_{\alpha}}$, with $h_{\alpha}$ and $\bar{h}_{\alpha}$ its holomorphic and antiholomorphic conformal dimensions \cite{CFT2}. 

The resulting $\bra{\phi_\beta^{\CFT}} V^{\CFT} \ket{\phi_{\alpha}^{\CFT}}$ for specific choices of profile functions $a(x)$ and $b(x)$ (described below), were compared in Figs. \ref{fig:time}-\ref{fig:scale} of the main text with matrix elements $\bra{\phi_\beta} V \ket{\phi_{\alpha}}$ numerically obtained on the spin chain, where $\ket{\phi_{\alpha}}$ and $V$ denote an energy/momentum eigenstate of the spin chain and a linear map implemented by a tensor network, respectively.

The generator $Q = Q_0 + iQ_1$ can be expanded in terms of the Fourier modes $H_n = L_n +\bar{L}_{-n}$ and $P_n=L_n -\bar{L}_{-n}$ of the hamiltonian and momentum densities $h(x)$ and $p(x)$, namely
\begin{equation}
Q = \frac{2\pi}{L}\sum_{n\in \mathbb{Z}} \left(a_n H_n + ib_n P_n\right)
\end{equation} 
Thus, we would like to compute the matrix elements $\bra{\phi_\beta} Q^n \ket{\phi_{\alpha}}$, were $Q^n$ is a sum of products of powers of $H_n$ and $P_n$. To do this computation, we express $Q^n$ as a sum of products of Virasoro generators such as $X \equiv \left(L_{n_1} \dots L_{n_r}\right) \left(\bar{L}_{\bar{n}_1} \dots \bar{L}_{\bar{n}_s} \right)$ for some integers $r$ and $s$ and apply the Virasoro algebra. As a first step, we express the states $|\phi_\alpha^{\CFT}\rangle$ in terms of Virasoro generators acting on primary states.

\subsection{Descendant states in terms of primary states}

To keep the notation relatively simple, given a primary state $\ket{h,\bar{h}}$ we consider a descendant state $\ket{\phi_{\alpha}^{\CFT}}$ that is obtained by applying a sequence $L_{n_1}L_{n_2} \cdots L_{n_q}$ of holomorphic Virasoro generators $L_n$,
\begin{equation} 
\ket{\phi_{\alpha}^{\CFT}} = \frac{1}{N_{\alpha}} L_{n_1}L_{n_2} \cdots L_{n_q} \ket{h,\bar{h}},
\end{equation}
where $n_i \leq -1$ and $N_{\alpha}$ is a normalization factor such that $\braket{\phi_{\alpha}^{\CFT}}{\phi_{\alpha}^{\CFT}}=1$. $N_{\alpha}$ can be expanded as
\begin{eqnarray}
N_{\alpha}^2 &=& \bra{h,\bar{h}}\left(L_{n_1}L_{n_2} \cdots L_{n_q}\right)^{\dagger} L_{n_1}L_{n_2} \cdots L_{n_q} \ket{h,\bar{h}} ~~~\\
&=& \bra{h,\bar{h}} \left(L_{n_q}^{\dagger} \cdots L_{n_2}^{\dagger} L_{n_1}^{\dagger} \right) L_{n_1}L_{n_2} \cdots L_{n_q} \ket{h,\bar{h}}~~~\\
&=& \bra{h,\bar{h}} \left(L_{-n_q} \cdots L_{-n_2} L_{-n_1} \right) L_{n_1}L_{n_2} \cdots L_{n_q} \ket{h,\bar{h}},~~~~~
\end{eqnarray}
where we used that $L_n^{\dagger} = L_{-n}$. A more general descendant state is descended from a primary state via both $L_n$ and $\overline{L}_{\bar{n}}$ generators:
\begin{equation}\label{eq:desc_state}
\ket{\phi_{\alpha}^{\CFT}} = \frac{1}{N_{\alpha}} \left(L_{n_1}L_{n_2} \cdots L_{n_q}\right)  \left(\bar{L}_{\bar{n}_1}\bar{L}_{\bar{n}_2} \cdots \bar{L}_{\bar{n}_q} \right)\ket{h,\bar{h}},
\end{equation}
with $n_i, \bar{n}_i \leq -1$, and its norm $N_{\alpha}$ can be expanded similarly.

\subsection{Matrix elements of virasoro generators}

Given a product of Virasoro generators $X$, we wish to compute
\begin{equation}
  X_{\alpha\beta} \equiv \langle \phi_\beta^{\CFT} | X | \phi_\alpha^{\CFT} \rangle.
\end{equation}
We first expand the states $|\phi_\alpha^{\CFT}\rangle$ and  $|\phi_\beta^{\CFT}\rangle$ in terms of their primary states and products of Virasoro generators as in the previous section. If the two primary states are different, thus belonging to different representations of the Virasoro algebra, $X_{\alpha\beta}=0$. If the primary states are the same, we have
\begin{equation}
  X_{\alpha\beta} \equiv \frac{1}{N_\alpha N_\beta}\langle h,\bar{h} | \left(L_{n_1} \dots L_{n_r}\right) \left(\bar{L}_{\bar{n}_1} \dots \bar{L}_{\bar{n}_s} \right) | h,\bar{h} \rangle,
\end{equation}
where we have also expanded $X$, resulting in an expectation value on the state $|h,\bar{h}\rangle$ of a long sequence of generators. We have also used the commutativity of $L_n$ with $\bar{L}_m$. To evaluate the expectation value, we apply the Virasoro commutators of Eq.~(\ref{eq:Virasoro}) to produce ``normal ordered'' terms, moving all generators with $n>0$ ($\bar{n}>0$) to the right and all generators with $n<0$ ($\bar{n}<0$) to the left. Since $L_n\bar{L}_{m}|h,\bar{h}\rangle = 0$ for $n>0$ or $m>0$, and $L_n^\dagger = L_{-n}$, $\bar{L}_n^\dagger = \bar{L}_{-n}$, these terms are zero and we are left with terms proportional to
\begin{equation}
  \langle h,\bar{h}|(L_0)^a(\bar{L}_0)^b|h,\bar{h}\rangle = (h)^a (\bar{h})^b,
\end{equation}
for integers $a,b \ge 0$, where we have used $L_0|h,\bar{h}\rangle=h|h,\bar{h}\rangle$ and $\bar{L}_0|h,\bar{h}\rangle=\bar{h}|h,\bar{h}\rangle$. We sum these terms to compute the final result.

The normalization factors $N_\alpha$ can be computed in the same way. For example, the norm-squared of the state $|\phi^{\CFT}\rangle = L_{-2}|h,\bar{h}\rangle$ is
\begin{align}
  N_\phi^2 &= \langle h,\bar{h} | L_2 L_{-2}  | h,\bar{h} \rangle \\
           &= \langle h,\bar{h} | 4L_0 + \frac{c}{2}  | h,\bar{h} \rangle\\
  &= 4h+\frac{c}{2},
\end{align}
where we applied the Virasoro commutator once. In particular, the norm of the state $|T^{\CFT}\rangle = L_{-2}|\mathbb{I}^{\CFT}\rangle$ is $N_T = \sqrt{c/2}$, since for the identity primary $h=\bar{h}=0$. An exemplary off-diagonal matrix element involving this state is
\begin{align}
&  \langle T^{\CFT} | \bar{L}_3 L_{-2}\bar{L}_{-3} |\mathbb{I}^{\CFT}\rangle \\
=~ &\frac{1}{N_T}\langle \mathbb{I}^{\CFT} | \bar{L}_3 \bar{L}_{-3} L_2 L_{-2} | \mathbb{I}^{\CFT} \rangle \\
=~ & N_T \langle \mathbb{I}^{\CFT} | \bar{L}_3 \bar{L}_{-3} | \mathbb{I}^{\CFT} \rangle \\
= ~& N_T \langle \mathbb{I}^{\CFT}| 6L_0 + 2c |\mathbb{I}^{\CFT} \rangle \\
= ~& \sqrt{2}c^{3/2}.
\end{align}

%Using this procedure, we analytically compute matrix elements of the conformal transformations we compare to the the tensor networks presented in the main text.

\subsection{Finite non-uniform euclidean evolution}

We wish to compute matrix elements of a conformal transformation equivalent to evolving non-uniformly in euclidean time. In other words, the generator should have the form from Eq.~(\ref{eq:Q0})
\begin{equation} 
Q_0 \equiv \int_0^{L} dx~a(x)~h(x), 
\end{equation}
with $a(x)$ a real-valued periodic function of $x$ so that $a(0) = a(L)$. Since we want to compare the action of the finite transformation $V^{\CFT}_0 \equiv e^{-Q_0}$ on eigenstates of the CFT Hamiltonian to the action of the tensor network of Fig.~\ref{fig:time}(b) on corresponding low-energy eigenstates of a critical spin chain, we guess a form for $a(x)$ based on the structure of the tensor network. We first take each lattice site to represent an interval of the continuum one lattice-spacing in length and centered on that site. Then we interpret each euclideon as evolving its interval by one unit of euclidean time, which is consistent with the action of the full transfer matrix $\mathcal{T}$. We then guess that a smoother evolves its interval by, on average, \emph{half} a unit of time, with the amount of evolution changing linearly from zero at one end of the interval associated with the smoother, to one unit at the other end. This leads to the profile function shown in Fig.~\ref{fig:time}(a), where the way the smoother is ``cut'' from the euclideon determines the direction of its sloping profile: See section \ref{sec:time_smoothers}.

The profile function, defined as a continuous function of $x \in [0,L)$, is
\begin{equation}
  a(x) = \begin{cases}
    x & 0 < x \le 1 \\
    1 & 1 < x \le L/4 \\
    -x + L/4 & L/4 < x \le L/4+1 \\
    0 & L/4+1 < x \le L/2 \\
    x - L/2 & L/2 < x \le L/2+1 \\
    1 & L/2+1 < x \le 3L/4 \\
    -x + 3L/4 & 3L/4 < x \le 3L/4+1 \\
    0 & 3L/4+1 < x \le L
  \end{cases},
\end{equation}
and has the Fourier coefficients
\begin{equation} \label{eq:Fourier_boxy}
  a_n = e^{in S \frac{2\pi}{N}} \frac{1}{4} \;\mathrm{sinc}\left(\frac{n}{4}\right) \;\mathrm{sinc}\left(\frac{n}{N}\right) \left(1 + e^{in \pi}\right),
\end{equation}
where the shift $S$ reflects the freedom of choosing the origin along the x-axis. The corresponding conformal generator is
\begin{equation} \label{eq:Q_boxy}
  Q_0 = \frac{2\pi}{N} \sum_{n=-N/2}^{N/2} a_n H_n,
\end{equation}
where we have summed over only those modes that can be distinguished on a lattice of $N$ sites ($N$ is assumed to be even). For the plot in Fig.~\ref{fig:time}(b), where $N=24$, we expand the exponential $V^{\CFT}_0 \equiv e^{-Q_0}$ to 4th order in $Q_0$ so that all the matrix elements shown are reasonably well converged.

Note that there are some reasons to expect mismatches between $V^{\CFT}_0$ and the euclideon tensor network of Fig.~\ref{fig:time}(a), even with an appropriate choice of $a(x)$. Perhaps most significantly, due to the smoothers, the tensor network is not Hermitian, unlike $V^{\CFT}_0$. This is the cause of the slight differences in magnitude between matrix elements of the tensor network and their Hermitian conjugate counterparts in the plot of Fig.~\ref{fig:time}(b). Furthermore, we neglect the effect of finite-size corrections to the eigenstates of the critical spin chain due to irrelevant terms present in the lattice model, as well as the effects of any relevant or irrelevant operator contributions to the smoothers, which are indeed present, as demonstrated by the observed matrix elements connecting different conformal towers.

\subsection{Finite non-uniform scale transformation}

An appropriate conformal transformation for comparison with the local scaling tensor network of Fig.~\ref{fig:scale}(b) is generated by a position-dependent translation
\begin{equation} 
Q_1 \equiv \int_0^{L} dx~b(x)~p(x), 
\end{equation}
as in Eq.~(\ref{eq:Q1}), where $b(x)$ is a real-valued, periodic function of $x \in [0,L)$. Unlike in the euclidean time case, however, the function $b(x)$ cannot simply be scaled to produce the position-dependent translation function associated with a \emph{finite} transformation. Indeed, $b(x)$ represents a position-dependent \emph{velocity} and a point $x(s_0)$ on the x-axis is translated by an infinitesimal transformation $e^{i\epsilon Q_1}$ as
\begin{equation}
  x(s_0+\epsilon) = x(s_0) + \epsilon b(x(s_0)).
\end{equation}
To find the final location of a point $x(s_0)$ under a finite transformation $e^{isQ_1}$, this equation must be integrated, for example numerically. By doing this for every starting point of interest $x_j$, the translation
\begin{equation}
  B(x_j) \equiv x_j(s-s_0) - x_j(s_0)
\end{equation}
these points experience under the finite transformation can be computed.

By inspecting the tensor network of Fig.~\ref{fig:scale}(b) we can guess at the appropriate form for $B(x)$. We consider the action of the network on a state at the top, with each lattice site representing an unit interval, centered on the site, of continuous $x$-axis. The network first performs fine-graining on sites $1$ and $8$. We thus assign a constant scale-factor of $2$ to the interval $(8,1]$, \emph{between} sites $1$ and $8$. Since we do not know exactly how the smoothers contribute, we leave the behavior in the intervals $(7.5,8]$ and $(1,1.5]$ undetermined. The network then performs a coarse-graining of sites $4$ and $5$, hence we assign the scale factor $1/2$ to the interval $(3.5,5.5]$. We further assume there is no scaling in the neighborhood of points $2$ and $7$, since these each mark the midpoint of a pair of smoothers and, by symmetry, the scale factor should pass through zero here.

To derive shifts, we must further pay attention to how the outgoing legs at the bottom of the network are arranged relative to the ingoing legs at the bottom. Although the unscaled sites $2$ and $7$ pass through the smoothers without being translated, translations are needed at the bottom to match up the outgoing legs with lattice sites. Hence site $2$ is translated by $1$ unit (to the right) and site $7$ is translated by $-1$ (to the left). By assuming, based on the symmetry of the network, that the points $4.5$ and $8.5$ are not translated, we can then derive the translation of points in the intervals $(8,1]$ and $(3.5,5.5]$ from the above scale factors (a constant scale factor corresponds to a constant derivative of the translation function). We use only the values in the stated regions and at the stated points in arriving at a possible generator, reflecting our lack of knowledge concerning the precise action of the smoothers. The resulting profile
\begin{align}
  B(x) = \begin{cases}
    1 & x = 2 \\
    -x/2 + 7/4 & 3.5 < x \le 5.5 \\
    -1 & x = 7 \\
    2x - 16 & 8 < x \le 9 \;(\equiv 1)
  \end{cases}
\end{align}
is shown in Fig.~\ref{fig:scale}(a).

To constrain the set of generating functions $b(x)$, we restrict to those described by just the first two odd Fourier modes, which is enough to reproduce the regions of constant scaling quite accurately. We may restrict to odd functions since we know there must be nodes at $x=4.5$ and $x=8.5$. While it is possible to fit these parameters to $B(x)$, we choose Fourier coefficients that are very close to the optimal ones, but which better match the numerical matrix elements of Fig.~\ref{fig:scale}(b):
\begin{align} \label{eq:Fourier_cosy}
  b_1 = e^{i S \frac{2\pi}{N}} \; 0.5, \quad b_3 = e^{i 2 S \frac{2\pi}{N}} \; 0.0275,
\end{align}
where $S$ is a shift determined by the location of the origin along the x-axis. The generator of the corresponding scale transformation is
\begin{equation}
  Q_1 = \frac{2\pi}{N}\left( b_1 P_1 + b_1^* P_{-1} + b_3 P_3 + b_3^* P_{-3} \right).
\end{equation}
For the plot in Fig.~\ref{fig:scale}(b), where $N=8$, we expand the exponential $V^{\CFT}_1 \equiv e^{-iQ_1}$ to 7th order so that the plotted matrix elements are reasonably well converged.
 
\subsection{Final remark on $V_0^{\CFT}$ and $V_1^{\CFT}$ } 
 
Above we have carefully specified two profiles $a(x)$ and $b(x)$ that led to generators $Q_{0}$ and $Q_1$ for the specific conformal transformations $V_0^{\CFT}$ and $V_1^{\CFT}$ acting on the the CFT that we used in Figs.~\ref{fig:time} and \ref{fig:scale} for comparison with the linear maps implemented with tensor networks. The match, both qualitative and quantitative, between matrix elements of the linear maps $V$ and $V^{\CFT}$ obtained on the lattice with tensor networks and on the CFT with conformal transformations is remarkable, confirming that the proposed tensor networks indeed implement lattice versions of non-uniform euclidean time evolution and rescaling.

We remark that although the profiles $a(x)$ and $b(x)$ were obtained above through a somewhat convoluted derivation that required making ad hoc decisions, slightly different choices of profile $a(x)$ and $b(x)$ were also tested and seen to lead to very similar $V^{\CFT}$ whose matrix elements continued to accurately match the matrix elements of the proposed lattice linear map $V$. 
 
\section{Euclideons, disentanglers, isometries, and smoothers} 

In this section we briefly review the well-established construction of the tensors called eucliodeons $e$, disentanglers $u$, and isometries $w$, and sketch how to build smoothers $e_{L}, e_{R}, u_{L}, u_{R}$. 

\subsection{Euclideons}

An \textit{euclideon}, denoted $e$, is a tensor that implements euclidean time evolution. In Refs.~\cite{TNR,TNRMERA,TNRscale} the tensor $e$ was instead denoted as $A$. Two possible ways of building an euclideon are (i) from a quantum spin chain Hamiltonian $H$ (see e.g.\ Supplemental Material in \cite{TNRMERA}) and (ii) from the Boltzmann weights of the statistical partition function of a classical two-dimensional lattice system (see e.g.\ Ref.~\cite{TNR}). We briefly review those constructions for completeness. 

Given a local quantum spin chain Hamiltonian $H$, the defining property of an euclideon $e$ for $H$ is that a periodic row $\mathcal{T}$ of $N$ euclideons should implement the euclidean time evolution $\exp(-\tau H)$ for one unit of euclidean time $\tau=1$. If the spin chain is described at low energies by a relativistic quantum field theory (QFT), then we normalize $H$ such that the ground state energy is zero in the thermodynamic limit ($N\rightarrow \infty$) and the speed of light is $1$ \cite{Ash}. When the relativistic QFT is in addition a conformal field theory (CFT), the above normalization of $H$ implies that the euclideons are (up to lattice effects) isotropic in the two-dimensional euclidean space-time. 
After normalizing the Hamiltonian $H$ one first builds a matrix product operator (MPO) for $\exp(-\delta \tau H)$ for small $\delta \tau \ll 1$, e.g.\ using a Suzuki-Trotter decomposition, and then multiplies together $1/\delta \tau$ copies of the resulting MPO while appropriately truncating the bond indices of the resulting product $\left(\exp (-\delta \tau H) \right)^{1/\delta \tau} = \exp(- \tau H)$. This results in a new MPO made of (roughly) isotropic euclideons $e$, see e.g.\ part A of the Supplemental Material of Ref.~\cite{TNRMERA} for further details.

The construction of euclideons is even simpler if the quantum spin chain relates to a two-dimensional statistical model, since in this case we can express an euclideon $e$ directly in terms of Boltzmann weights. For instance, for the critical quantum Ising spin chain $H = -\sum_l \sigma^{x}_l\sigma^{x}_{l+1} - \sigma_l^{z}$, we can build euclideons using the Boltzmann weights $e^{\sigma_i\sigma_j/T}$ of the statistical partition function of the (isotropic) critical 2d classical Ising model, namely
\begin{equation} \label{eq:A}
e_{ijkl} \equiv e^{\left(\sigma_i\sigma_j + \sigma_j\sigma_k + \sigma_k\sigma_l + \sigma_l\sigma_i \right)/T},
\end{equation}
where $\sigma_i = \pm 1$ labels the two possible values of a classical Ising spin on site $i$ of a two-dimensional lattice, see e.g.\ Ref.~\cite{TNR} for further details.

\subsection{Euclidean time smoothers}
\label{sec:time_smoothers}

\begin{figure}
\includegraphics[width=7cm]{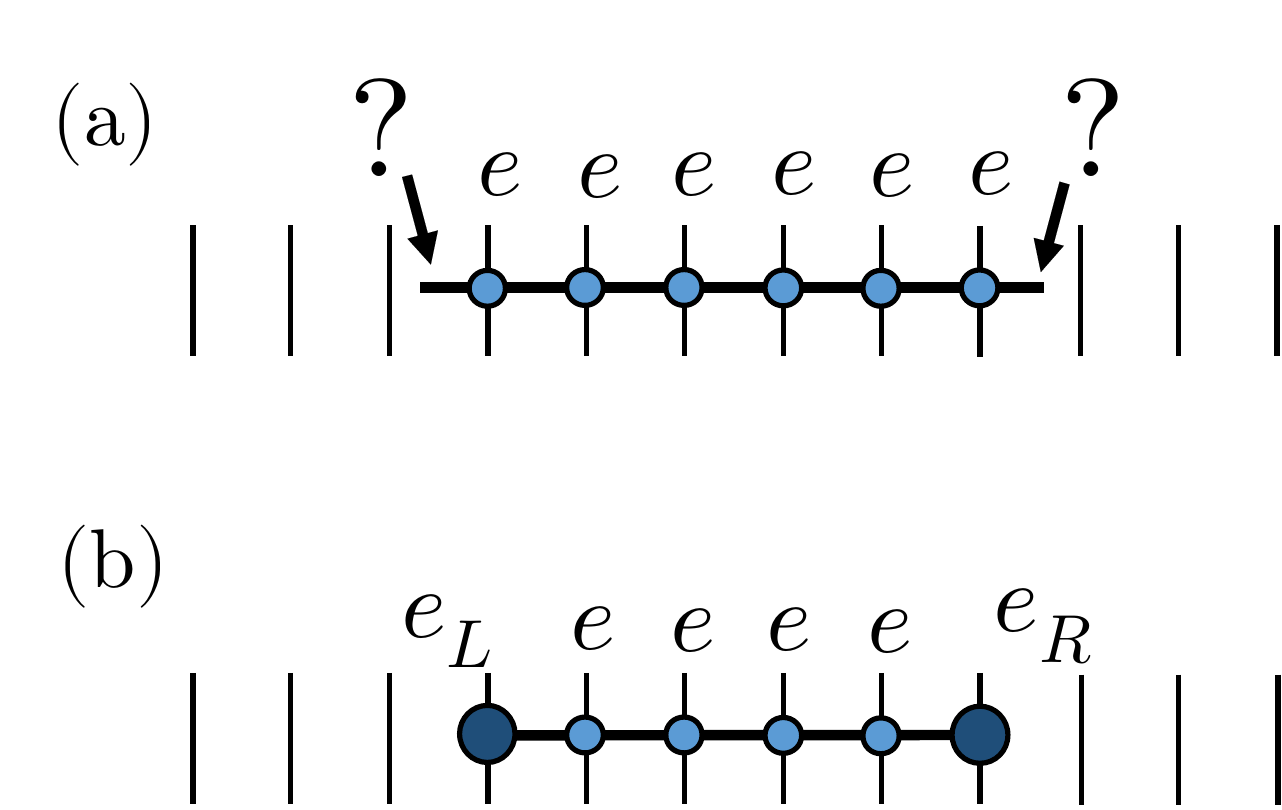}
\caption{
(a) A truncated row of euclideons $e$ has a dangling bond index at each end. As a result, it does not define a linear map in the Hilbert space of the spin chain.
(b) Left and right smoother $e_L$ and $e_R$ are special tensors placed at the two ends of the truncated row of euclideons in order to eliminate the dangling bond indexes, producing a linear map in the spin chain. 
\label{fig:timesmoother0} 
}
\end{figure}

We can use a truncated row of euclideons $e$ on a region of a quantum spin chain in order to apply an euclidean time evolution that only evolves the spins on that region. However, at both ends of the truncated row of euclideons we must place special tensors that we call \textit{smoothers} $e_L$ and $e_R$, see Fig.~\ref{fig:timesmoother0}. The purpose of each smoother is to smooth out lattice effects that occur at the ends of the truncated row of euclideons. Without smoothers there would be an open bond index at each end. Thus a first role of the smoothers, which only have one bond index (as opposed to the two bond indices of regular euclideons) is to eliminate open bond indices. However, not any tensor with a single bond index will do. Indeed, a generic choice of coefficients inside the smoother will result in a transformation that is not the intended conformal transformation (non-uniform euclidean time evolution). As a matter of fact, the resulting transformation is generically not even approximately diagonal in the conformal towers and therefore does not correspond to any conformal transformation.

\begin{figure}
\includegraphics[width=7cm]{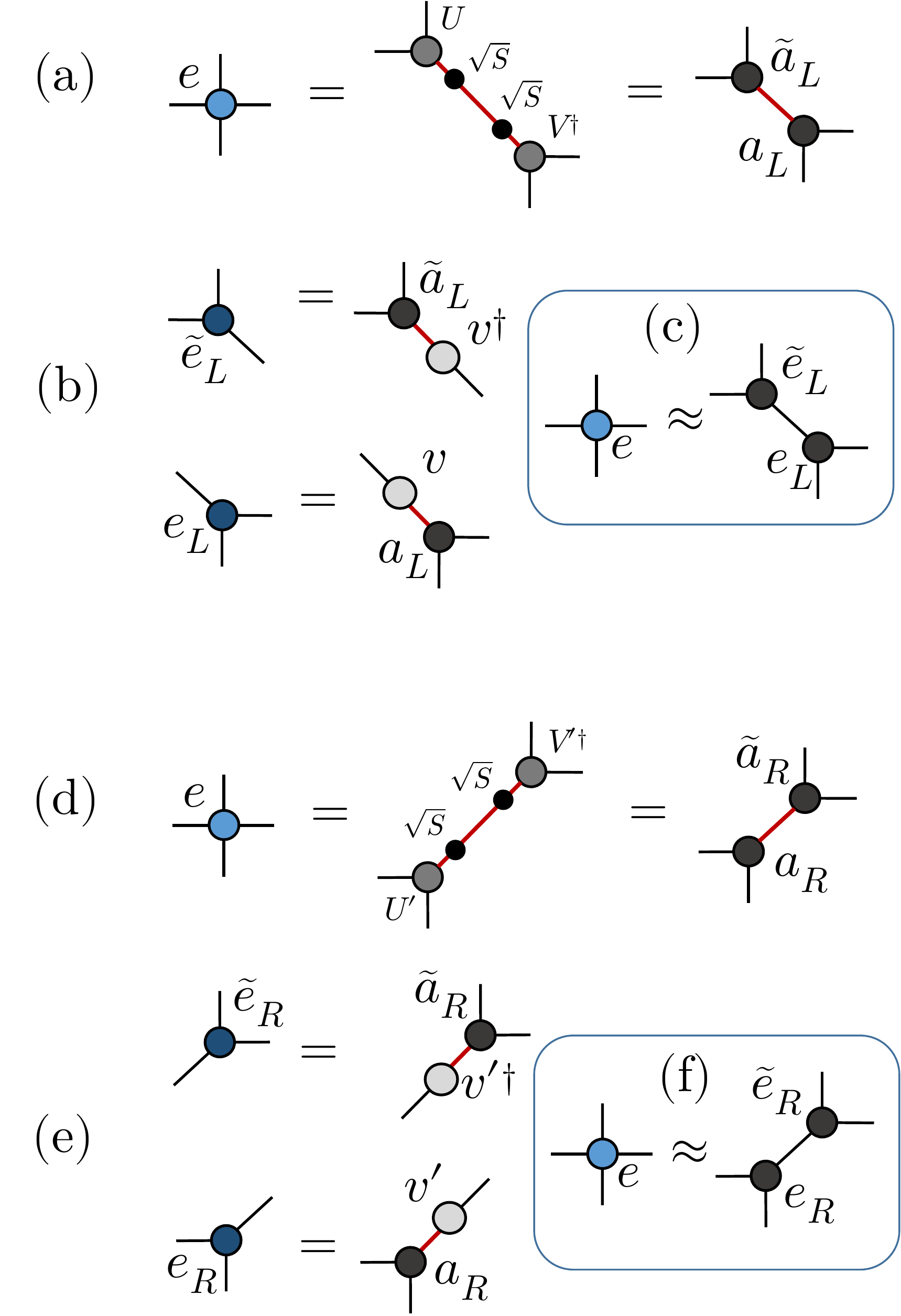}
\caption{
Computation of smoothers $e_L$ and $e_R$ starting from an euclideon $e$.
(a) In order to create a smoother $e_L$, first we temporarily regard an euclideon $e$ as a matrix $M$ (by joining its indexes into two pairs) and then compute its singular value decomposition $M= USV^{\dagger}$, where $U$ and $V$ are unitary matrices and $S$ is a diagonal matrix with the singular values of $M$ in its diagonal. Then we create tensors $a_L = \sqrt{S}V^{\dagger}$ and $\tilde{a}_L \equiv U\sqrt{S}$, where $\sqrt{S}$ is a diagonal matrix with the square root of the singular values in the diagonal. Tensor $a_L$, which is a precursor of the smoother $e_L$, has two indices (the ones which originally belonged to the euclideon $e$, depicted in black) of dimension $d$ and one index (the one coming from the singular value decomposition, depicted in red) of dimension $d^2$. 
(b) The smoother $e_L$ is then obtained by multiplying $a_L$ by an isometry $v$ of size $d\times d^2$ that maps the $d^2$-dimensional (red) index into a $d$-dimensional (black) index, thus implementing a truncation of the former. The variational parameters in $v$ correspond to a choice of truncation basis and are determined as indicated in Fig.~\ref{fig:timesmoother2}. 
(c) Notice that the product of $\tilde{e}_L$ (obtained analogously to $e_L$) and $e_L$ amounts to the original euclideon $e$, up to effects due to the truncation implemented through the isometry $v$.
(d)-(f) The smoother $e_R$ is created analogously. 
\label{fig:timesmoother1} 
}
\end{figure}

\begin{figure}
\includegraphics[width=8.5cm]{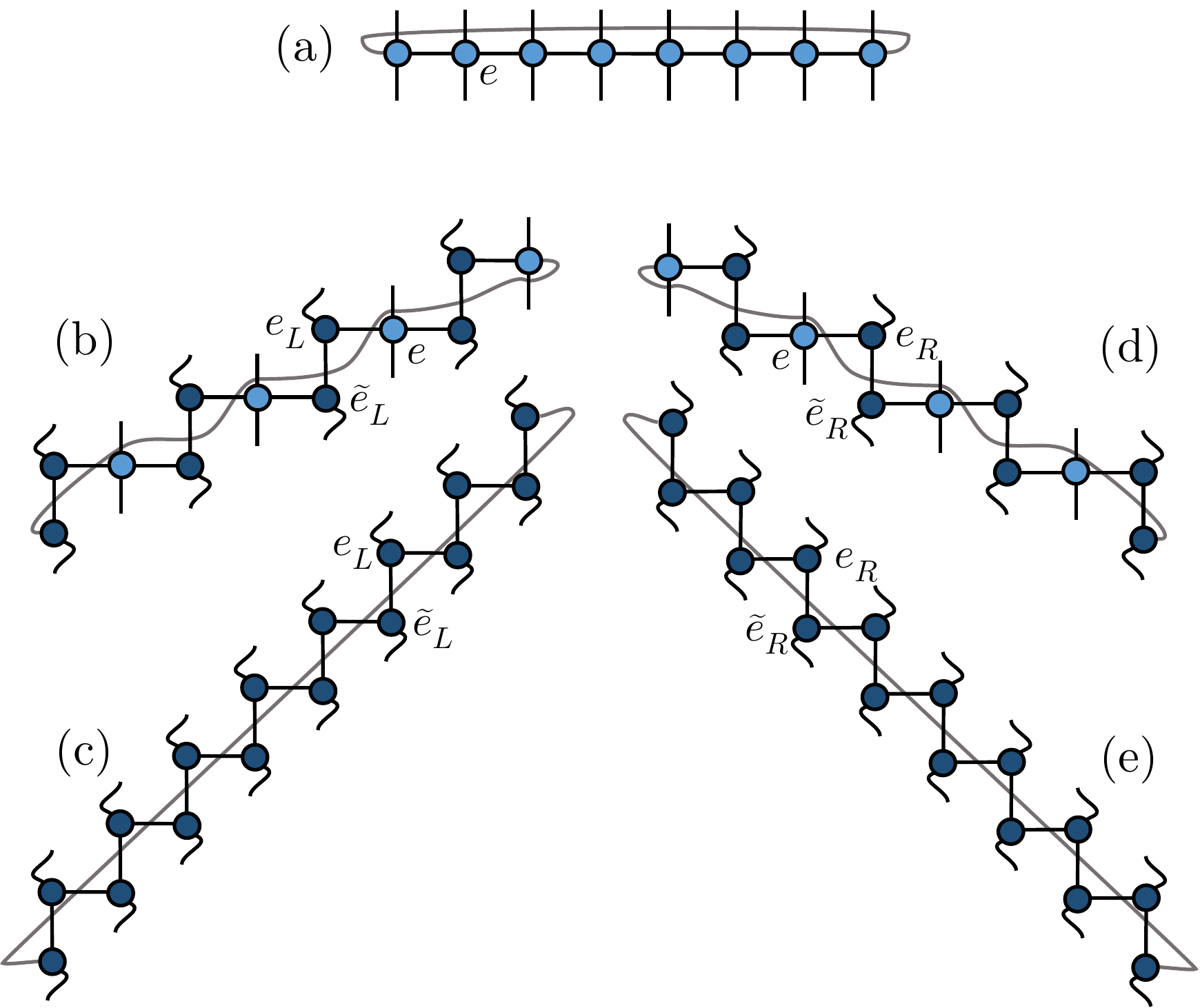}
\caption{
In order to determine the variational parameters in the smoother $e_L$, which are contained in the isometry $v$ of Fig.~\ref{fig:timesmoother1}(b), we require that the dominant eigenvector of the transfer matrix $\mathcal{T}$ (a), corresponding to the ground state on the circle, has maximal overlap with the dominant eigenvector of other transfer matrices, such as those in (b) and (c), which can be seen to become close to the ground state on the circle. The variational parameters in the smoother $e_R$ are determined analogously using the dominant eigenvectors of transfer matrices $(d)$ and $(e)$.
\label{fig:timesmoother2} 
}
\end{figure}

In practice we build the smoothers through a two-step procedure, see Fig.~\ref{fig:timesmoother1}. First we split an euclideon diagonally using a singular value decomposition (SVD). Then we multiply it by an isometry, that we determine by demanding that it optimally connects to the physical indices of the original euclideon $e$, in the following sense. We build a \textit{diagonal transfer matrix} $M'$ made of euclideons and smoothers and demand that its dominant eigenvector has maximal overlap with the dominant eigenvector of the horizontal transfer matrix $M$ made of euclideons. The resulting smoothers, when placed at the ends of a truncated row of euclideons, are then seen to indeed produce linear maps that not only act (approximately) diagonally in the conformal towers (as any conformal transformation does) but whose matrix elements between low energy states accurately correspond to the intended non-uniform euclidean time evolution, as seen in the example of Fig.~\ref{fig:time} in the main text.

\subsection{Optimized disentanglers and isometries} 
 
Disentanglers $u$ and isometries $w$, the tensors in the multi-scale entanglement renormalization ansatz (MERA), are in general full of variational parameters constrained in such a way that the tensors are unitary/isometric. In the context of this work, however, by a disentangler $u$ and an isometry $w$ we refer exclusively to such tensors after the variational parameters have been chosen so that the MERA represents the ground state of the quantum spin Hamiltonian $H$. 

This optimization can be carried out variationally using iterative energy minimization algorithms \cite{siMERA}, as we did here. Alternatively, one can extract optimized disentanglers and isometries from the tensor network renormalization (TNR) algorithm \cite{TNRMERA}, which manipulates a network of euclideons.

\subsection{Scale smoothers}

We can use a truncated double layer of optimized disentanglers and isometries on a region of a quantum spin chain in order to apply a non-uniform scale transformation that only rescales the spins in that region. Again, at both ends of the truncated layer of disentanglers and isometries we must place special tensors called smoothers $u_L$ and $u_R$. One first reason for including these smoothers is that otherwise there is a mismatch in index connectivity, since the lower indices of an isometry $w$ must connect with the upper indices of a disentangler $u$ (by design of the MERA). However, a generic choice of coefficients inside the scale smoothers $u_L$ and $u_R$ (even when compatible with the unitary/isometric character of these tensors) will produce a linear map that does not correspond to a conformal transformation. 

\begin{figure}
\includegraphics[width=8.5cm]{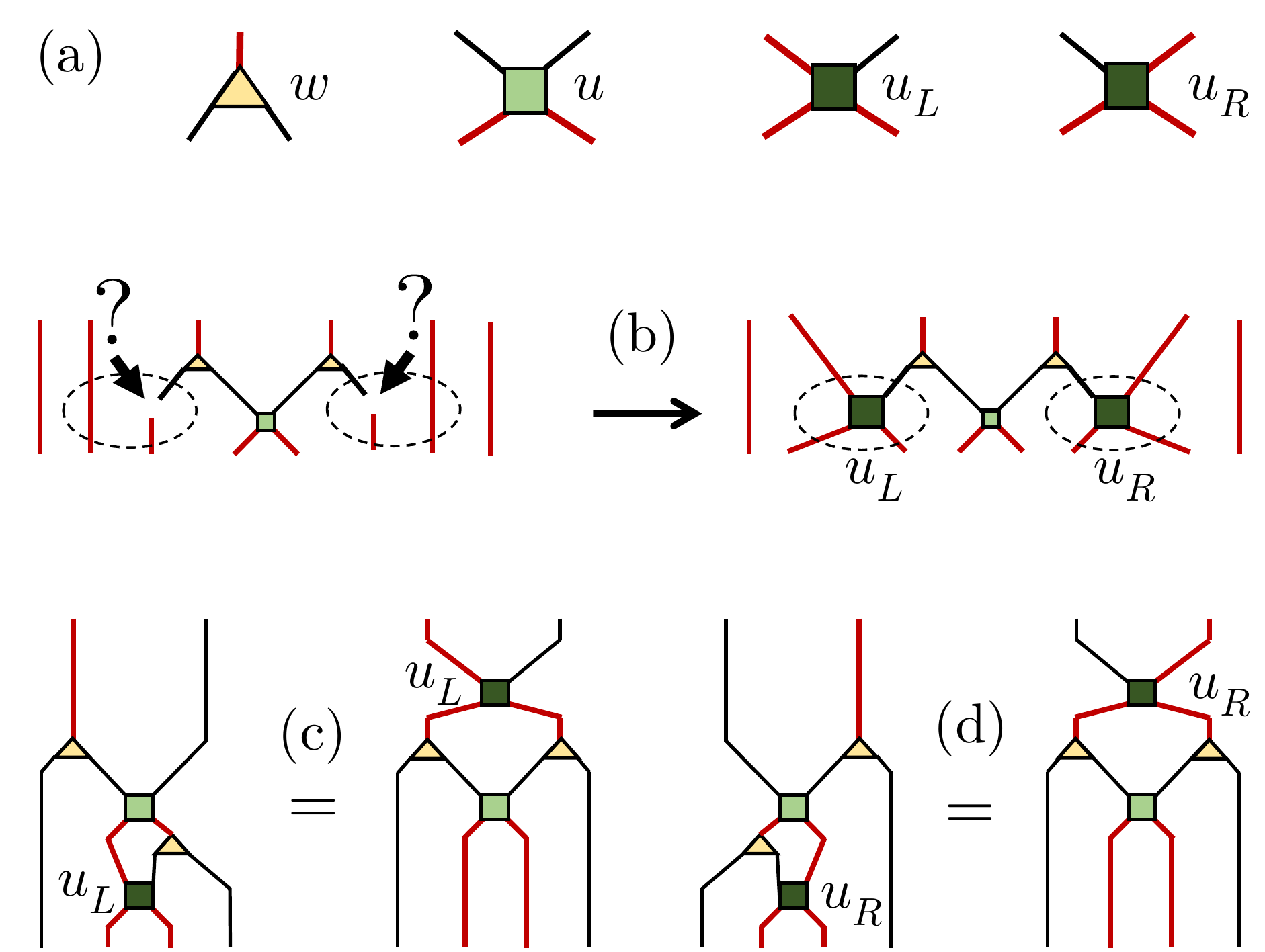}
\caption{
(a) Isometry $w$, disentangler $u$, and smoothers $u_L$ and $u_R$, with two types of indices: in red, the top index of an isometry $w$, which connects with the bottom indices of a disentangler $u$; in black, the bottom indices of an isometry $w$, which connect with the top indices of a disentangler $u$. Smoothers map a pair of red and black top indices into a pair of bottom red indices.
(b) At each end of a truncated double layer of MERA tensors $\mathcal{W}$ there is a dangling black index that cannot be directly connected to a red index. Scale smoothers $u_L$ and $u_R$ can map these dangling black indexes, together with an adjacent red index, into a pair of red indexes. 
(c) The left scale smoother $u_L$ is determined variationally by demanding that this tensor network equality be fulfilled (approximately, but as accurately as possible).
(d) Similarly, the right scale smoother $u_R$ is determined variationally by demanding that this other tensor network equality be approximately fulfilled.
\label{fig:scalesmoother1} 
}
\end{figure}

In practice, we build the smoothers $u_L$ and $u_R$ by demanding that the equality in Fig.~\ref{fig:scalesmoother1} be (approximately) fulfilled. These equalities ensure that the scale smoothers properly connect upper and lower indices. The resulting smoothers $u_L$ and $u_R$, when placed at the ends of a truncated double layer of optimized disentanglers and isometries, indeed produce a linear map that not only acts (approximately) diagonally in the conformal towers but whose matrix elements between low energy states accurately correspond to the intended non-uniform scale transformation.

\end{document}